\documentclass[twocolumn,showpacs,aps,prl,superscriptaddress]{revtex4}
\usepackage{graphicx}
\usepackage{dcolumn}
\usepackage{amsmath}
\usepackage{epsfig}
\usepackage{multirow}

\input babarsym

\newcommand{\BABARPubYear}    {04}
\newcommand{\BABARPubNumber}  {33}

\newcommand{\SLACPubNumber} {10629}

\def\Dsp      {\ensuremath{D^+_s}\xspace}
\def\Dsm      {\ensuremath{D^-_s}\xspace}
\def\Dsr      {\ensuremath{D_s}\xspace}
\def\Dspm     {\ensuremath{D^\pm_s}\xspace}

\def\Lambdabar {\kern 0.2em\overline{\kern -0.2em \Lambda}{}\xspace}
\def\fbar {\ensuremath{\kern 0.2em\overline{\kern -0.2em f}{}}\xspace}

\def\Lcp {\ensuremath{\Lambda^+_c}\xspace}
\def\Lcm {\ensuremath{\Lambdabar^-_c}\xspace}
\def\Lc  {\ensuremath{\Lambda_c}\xspace}

\def\Brec {\ensuremath{\B_{\mathrm reco}}\xspace}

\def\figurebox#1#2#3{%
    \def\arg{#3}%
    \ifx\arg\empty
    {\hfill\vbox{\hsize#2\hrule\hbox to #2{\vrule\hfill\vbox to #1{\hsize#2\vfill}\vrule}\hrule}\hfill}%
    \else
    {\hfill\epsfbox{#3}\hfill}%
    \fi}

\begin{document}

\preprint{\babar-PUB-\BABARPubYear/\BABARPubNumber}
\preprint{SLAC-PUB-\SLACPubNumber}

\begin{flushleft}
\babar-PUB-\BABARPubYear/\BABARPubNumber\\
SLAC-PUB-\SLACPubNumber\\
\end{flushleft}

\title{
{\large \bf Measurement of the Branching Fractions for Inclusive \boldmath \Bm and \Bzb Decays to Flavor-tagged $D$, \Dsr and \Lc } }

%
\author{B.~Aubert}
\author{R.~Barate}
\author{D.~Boutigny}
\author{F.~Couderc}
\author{J.-M.~Gaillard}
\author{A.~Hicheur}
\author{Y.~Karyotakis}
\author{J.~P.~Lees}
\author{V.~Tisserand}
\author{A.~Zghiche}
\affiliation{Laboratoire de Physique des Particules, F-74941 Annecy-le-Vieux, France }
% ...
\collaboration{The \babar\ Collaboration}
\noaffiliation

\date{\today}

\begin{abstract}
We report on the inclusive branching fractions of \Bm and of \Bzb mesons decaying to 
${\Dz \X}$, ${\Dzb \X}$, ${\Dp \X}$, ${\Dm \X}$, ${\Dsp \X}$, ${\Dsm \X}$, ${\Lcp \X}$, ${\Lcm \X}$, 
based on a sample of 88.9 million $\BB$ events recorded with the \babar\ detector at the \Y4S resonance. 
Events are selected by completely reconstructing one $\B$ and searching for a 
reconstructed charmed particle in the rest of the event. 
We measure the number of charmed and of anti-charmed particles per $B$ decay and 
derive the total charm yield per \Bm decay, 
$n_\c^- = 1.313 \pm 0.037 \pm 0.062 ^{+0.063}_{-0.042} $, 
and per \Bzb decay, 
$n_\c^0 = 1.276 \pm 0.062 \pm 0.058 ^{+0.066}_{-0.046}$
where the first uncertainty is statistical, 
the second is systematic, and the third reflects the charm branching-fraction 
uncertainties.
\end{abstract}


\maketitle

The dominant process for the decay of a \b quark is $\b\to\c \W^{*-}$~\cite{chconj}, 
resulting in a (flavor) correlated \c quark and a virtual \W. In the decay of the \W, 
the production of a $\ubar d$ or a $\cbar s$ pair are both Cabibbo-allowed and should be equal, 
the latter being only suppressed by a phase-space factor. 
The first process dominates hadronic \b decays, while
the second can be easily distinguished as it will produce a (flavor) anti-correlated \cbar quark. 
Experimentally, correlated and anti-correlated charm production can be investigated 
through the measurement of the inclusive \B-decay rates to flavor-tagged charmed mesons or
baryons. Current measurements~\cite{cleoinc,delphiinc,cleolc} of these rates have 
statistically limited precision and do not distinguish among the different \B 
parent states.

Most of the charged and neutral $D$ mesons produced in \Bb decays come from correlated production 
$\Bb \to D X$. However, a significant number of $\Bb \to \Db X$ decays are expected through 
$\b\to\c\cbar\s$ transitions, such as $\Bb \to D^{(*)}\Db^{(*)} \Kb^{(*)}(n\pi)$. 
Although the branching fractions of the 3-body decays $\Bb\to D^{(*)} \Db^{(*)} \Kb$ 
have been measured~\cite{alephddk,babarddk}, it is not clear whether they saturate 
$\Bb\to \Db X$ transitions. It is therefore important to improve the precision on 
the branching fraction $\BR(\Bb\to \Db X)$.

By contrast, the anti-correlated \Dsm production ${\Bb\to \Dsm D (n\pi)}$ is expected 
to dominate \Bb decays to \Dsr mesons, since 
correlated production needs an extra \ssbar pair created from the vacuum to give ${\Bb\to \Dsp \Km (n\pi)}$. 
There is no prior published measurement of $\BR(\Bb\to \Dsp X)$.

All strangeless charmed baryons decay to \Lc. Correlated \Lc are produced in decays like 
$\Bm\to\Lcp \antiproton \pim(\pi)$, while anti-correlated $\Lcm$ should originate from 
$\Bm\to\Xi_c\Lcm(\pi)$. Another possibility is $\Bm\to \Lcp \Lcm \Km$, 
the baryonic analogue of the $D \Db \kaon$ decay.  
The rates for $\Xi_c$ production in \B decays~\cite{cleoxicr} are unknown, 
because there is no  absolute measurement of $\Xi_c$ decay branching fractions.

This analysis uses \upsbb events in which
either a \Bp or a \Bz meson  (hereafter denoted \Brec )
decays into a hadronic final state and is fully reconstructed.
We then reconstruct $D$, \Dsr and \Lc from the recoiling \Bm (\Bzb) meson 
and compare the flavor of the charm hadron with that 
of the \Brec, 
thus allowing separate measurements of the 
\Bm\ (\Bzb)\ \to\Dz\X, \Dp\X, \Dsp\X, \Lcp\X and 
\Bm\ (\Bzb)\ \to\Dzb \X, \Dm\X, \Dsm\X, \Lcm\X branching fractions. 
We extract ${\BR(\Bm\to \Lcp \Lcm \Km)}$ from the missing-mass spectra 
of the $\Lcp \Km$ or $\Lcm \Km$ systems recoiling against the \Brec. 
We can then evaluate indirectly 
$\BR(\Bm\to \Xi_c \X) = \BR(\Bm\to \Lcm \X) - \BR(\Bm\to \Lcp \Lcm \Km)$ 
and compute the average number of charm (anti-charm) particles per \Bm decay, 
$N_\c^- $ ($N_{\cbar}^-$):
\begin{eqnarray}
    N_\c^-      & = & \sum_{\X_c}    \BR(\Bm\to \X_c X),\label{eq:nc} \\
    N_{\cbar}^- & = & \sum_{\Xb_c} \BR(\Bm\to \Xb_c X), \label{eq:ncbar}
\end{eqnarray}
where the sum is performed over $\X_c=$ \Dp, \Dz, \Dsp, \Lcp, $\Xi_c$, $(\ccbar)$ 
or $\Xb_c=$ \Dm, \Dzb, \Dsm, \Lcm, $(\ccbar)$, and $(\ccbar)$ 
refers to all charmonium states collectively.
We neglect $\Xibar_c$ production, 
as it requires both a $\cbar\s$ and an \ssbar pair in the decay to give $\Xibar_c\Omega_c$. 
We can sum $N_\c^-$ and $N_{\cbar}^-$ to obtain the average number of 
charm plus anti-charm quarks per \Bm decay, $n_\c^- = N_\c^- + N_{\cbar}^-$ 
(and similarly for \Bzb decays). 
In addition to the theoretical interest~\cite{ncbagan, ncbuchalla, ncneubert}, 
the fact that anti-correlated charmed particles are a background for many studies 
also motivates a more precise measurement of their production rates in \B decays.

The measurements presented here are based on a sample of 
$88.9$ million \BB pairs ($81.9~\invfb$) recorded at the \Y4S resonance 
with the \babar\ detector at the PEP-II asymmetric-energy \B-meson factory at SLAC. 
The \babar\ detector is described in detail elsewhere~\cite{det}. 
Charged-particle trajectories are measured by a 5-layer double-sided silicon vertex tracker 
and a 40-layer drift chamber, both operating in a 1.5-T solenoidal magnetic field. 
Charged-particle identification is provided by the average energy loss (\dedx) 
in the tracking devices and by an internally reflecting ring-imaging Cherenkov detector. 
Photons are detected by a CsI(Tl) electromagnetic calorimeter. 
We use Monte Carlo simulations of the \babar\ detector based on GEANT4~\cite{GEANT} 
to optimize selection criteria and determine selection efficiencies.

We reconstruct \Bp and \Bz decays (\Brec) in the modes 
$\Bp\to \Db^{(*)0}\pip$, $\Db^{(*)0}\rho^+$, $\Db^{(*)0}a_1^+$ and 
$\Bz\to D^{(*)-}\pip$, $D^{(*)-}\rho^+$, $D^{(*)-}a_1^+$. 
\Dzb candidates are reconstructed in the $\Kp\pim$, $\Kp\pim\piz$, 
$\Kp\pim\pip\pim$ and $\KS\pip\pim$($\KS\to\pip\pim$) decay channels, 
while \Dm are reconstructed in the $\Kp\pim\pim$ and $\KS\pim$ modes. 
\Dstar candidates are reconstructed in the $D^{*-}\to \Dzb\pim$ and 
$\Dstarzb\to \Dzb\piz$, $\Dzb\g$ decay modes. 
The first kinematic variable used to identify fully reconstructed \B decays
is the beam-energy substituted mass,
$\mes =\sqrt{ (s/2 + {\bf p}_{i}\cdot{\bf p}_{B})^{2}/E_{i}^{2}-{\bf p}^{2}_{B}}$,
where ${\bf p}_{B}$ is the \Brec momentum and  
$(E_{i},{\bf p}_{i})$ is the four-momentum of the initial \epem system,
both measured in the laboratory frame.
The invariant mass of the initial \epem system is $\sqrt{s}$. 
The second variable is $\DeltaE = E^*_B-\sqrt{s}/2$, 
where $E^*_B$ is the \Brec candidate energy in the center-of-mass frame.
We require $|\DeltaE|<n\,\sigma_{\Delta E}$ with $n=2$ or $3$, 
depending on the decay mode, and using the measured resolution 
$\sigma_{\Delta E}$ for each decay mode.

\begin{figure}[!h]
\begin{center}
\includegraphics[width=0.90\linewidth]{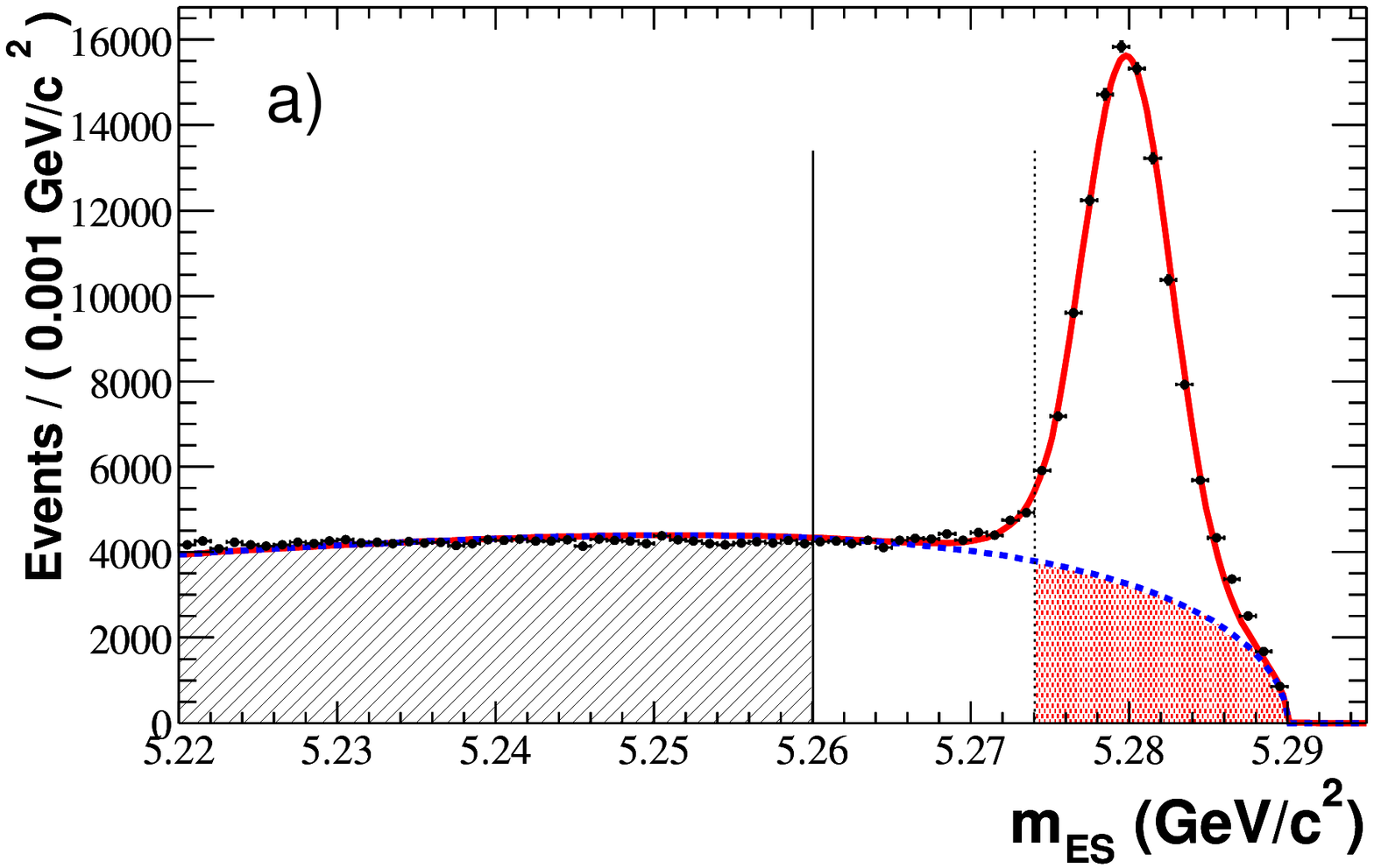}
\includegraphics[width=0.90\linewidth]{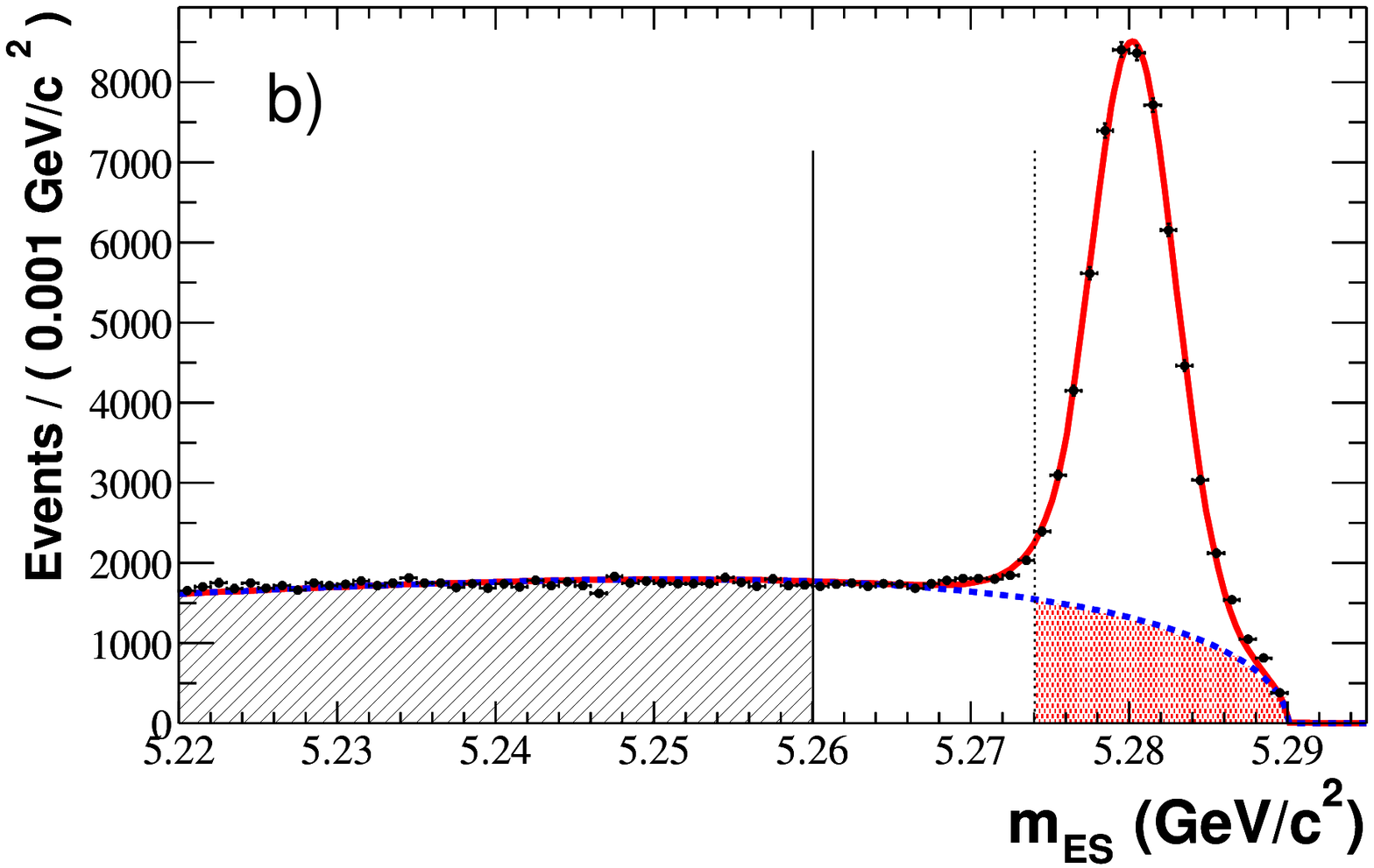}
\caption{\mes spectra of reconstructed (a) \Bp and (b) \Bz candidates. 
The full vertical line shows the upper limit of the background
control region (hatched), the dotted vertical line the lower limit of the $B$ signal region.
The crossed area shows the background under the \B signal. 
The solid curve is the sum of the fitted signal and background, 
the dashed curve is the background component only.} \label{fig:mes}
\end{center}
\end{figure}

In the \mes spectra (Fig.~\ref{fig:mes}), we define  a signal region with 
$5.274<\mes<5.290~\gevcc$ and a background control region with 
$5.220<\mes<5.260~\gevcc$. For each of the \B-decay modes, 
the combinatorial background in the signal region is derived from a fit 
to the \mes distribution that uses an empirical phase-space 
threshold function~\cite{fargus} for the background, 
together with a signal function~\cite{fcrystalball} peaked at the \B meson mass. 
The numbers of reconstructed \Bp  and \Bz candidates, 
$N_{\Bp}=85840\pm 1910\ ({\rm syst.})$ and $N_{\Bz}=48322 \pm 590\ ({\rm syst.})$, 
are then obtained by subtracting this background 
from the total number of events found in the signal region.
These measured \B meson yields provide the normalization of all 
branching fraction measurements reported below.
The systematic uncertainties quoted above are computed 
by varying the boundaries of the signal and background regions, 
and by comparing the shapes of the threshold function~\cite{fargus} 
in the data and in the simulation.

The contamination of \Bz events in the \Bp signal induces a background 
which peaks near the \B mass. From the Monte Carlo simulation, 
the fraction of \Bz events
in the reconstructed \Bp signal sample
is found to be $c_0=0.034$, and 
the fraction of \Bp events 
in the reconstructed \Bz signal sample to be $c_+=0.019$.
A $100$~\% systematic uncertainty is conservatively assigned to these numbers
but they will have a small effect on the final results.

We now turn to the analysis of inclusive $D$, \Dsr and \Lc production 
in the decays of the $\Bb$'s that recoil against the reconstructed $B$. 
Charmed particles $\X_c$ (correlated production) are distinguished 
from anti-charmed particles $\Xb_c$ (anti-correlated production). 
They are reconstructed from charged tracks that do not belong to the \Brec. 
The decay modes considered are listed in Table~\ref{table:t1}.

\begin{table*}[!htb]
  \centering
  \caption{Charmed-particle signal yields and $\B$ branching fractions per decay mode. 
  The first uncertainty is statistical, the second is systematic 
  (but does not include the charm branching fraction uncertainties).
}\label{table:t1}
\begin{tabular}{l@{$\to$}l r@{$\pm$}l r@{$\pm$}c@{$\pm$}l  r@{$\pm$}l r@{$\pm$}c@{$\pm$}l  r@{$\pm$}l r@{$\pm$}c@{$\pm$}l  r@{$\pm$}l r@{$\pm$}c@{$\pm$}l}
  \hline\hline
  \multicolumn{22}{c}{}\\
  \multicolumn{2}{c}{$\X_c$ decay mode} & 
  \multicolumn{5}{c}{$\Bm \to \X_c \X$} &
  \multicolumn{5}{c}{$\Bm \to \Xb_c\X$} & 
  \multicolumn{5}{c}{$\Bzb\to \X_c \X$} &
  \multicolumn{5}{c}{$\Bzb\to \Xb_c\X$}\\
  \multicolumn{2}{c}{  }                & 
  \multicolumn{2}{c}{yield} & \multicolumn{3}{c}{$\BR$(\%)}&
  \multicolumn{2}{c}{yield} & \multicolumn{3}{c}{$\BR$(\%)}&
  \multicolumn{2}{c}{yield} & \multicolumn{3}{c}{$\BR$(\%)}&
  \multicolumn{2}{c}{yield} & \multicolumn{3}{c}{$\BR$(\%)} \\ \hline

  $\Dz$  & $\Km\pip$          & $1273$ & $42$& $79.2$ & $2.6$ & $3.9$ & $160$ & $16$ &  $9.3$ & $1.0$ & $0.5$
                              & $397$ & $24$ & $50.3$ & $3.4$ & $2.4$ & $139$ & $14$ & $ 7.3$ & $2.2$ & $0.5$\\
         & $\Km\pip\pim\pip$  & $998$ & $65$ & $80.6$ & $5.3$ & $7.5$ & $173$ & $30$ & $13.4$ & $2.4$ & $1.3$
                              & $332$ & $36$ & $56.2$ & $6.8$ & $5.4$ & $ 83$ & $23$ & $ 1.8$ & $4.4$ & $0.5$\\ \hline
  $\Dp$ & $\Km\pip\pip$       & $262$ & $29$ &  $9.8$ & $1.2$ & $1.2$ &  $98$ & $20$ & $ 3.8$ & $0.9$ & $0.4$
                              & $452$ & $31$ & $39.7$ & $3.0$ & $2.8$ & $125$ & $18$ & $ 2.3$ & $1.8$ & $0.3$\\ \hline
 $\Dsp$ & $\phi\pip$          & $11$ & $5$   & $2.2$ & $1.1$ & $0.3$  & $82$ & $11$  & $16.5$ & $2.3$ & $1.7$
                              & $24$ & $ 6$  & $8.3$ & $2.8$ & $0.8$  & $ 28$ & $ 6$ & $ 9.9$ & $2.9$ & $1.0$\\
       & $\Kstarzb\Kp$        &  $0$ & $3$   & $0.0$ & $1.1$ & $0.2$  & $55$ & $11$  & $18.0$ & $3.5$ & $1.7$
                              &  $3$ & $4$   & $0.0$ & $2.8$ & $0.1$  & $14$ & $ 5$  & $ 9.9$ & $4.1$ & $1.2$\\
          & $\KS \Kp$         &  $0$ & $3$   & $0.0$ & $0.9$ & $0.2$  & $31$ & $ 9$  & $ 9.2$ & $2.7$ & $0.8$
                              & $12$ & $5$   & $5.0$ & $3.4$ & $0.4$  & $ 23$ & $ 6$ & $13.3$ & $4.3$ & $1.0$\\
  \hline
 $\Lcp$ & $\proton\Km\pip$    & $41$ & $9$   & $3.5$ & $0.8$ & $0.3$  &  $33$ & $9$  & $2.9$ & $0.8$ & $0.3$
                              & $28$ & $8$   & $4.9$ & $1.7$ & $0.4$  & $16$ & $ 6$  & $2.0$ & $1.2$ & $0.2$\\
  \hline\hline
\end{tabular}\end{table*}
\begin{figure}[!ht]
\begin{center}
\includegraphics[width=0.98\linewidth,height=0.42\linewidth]{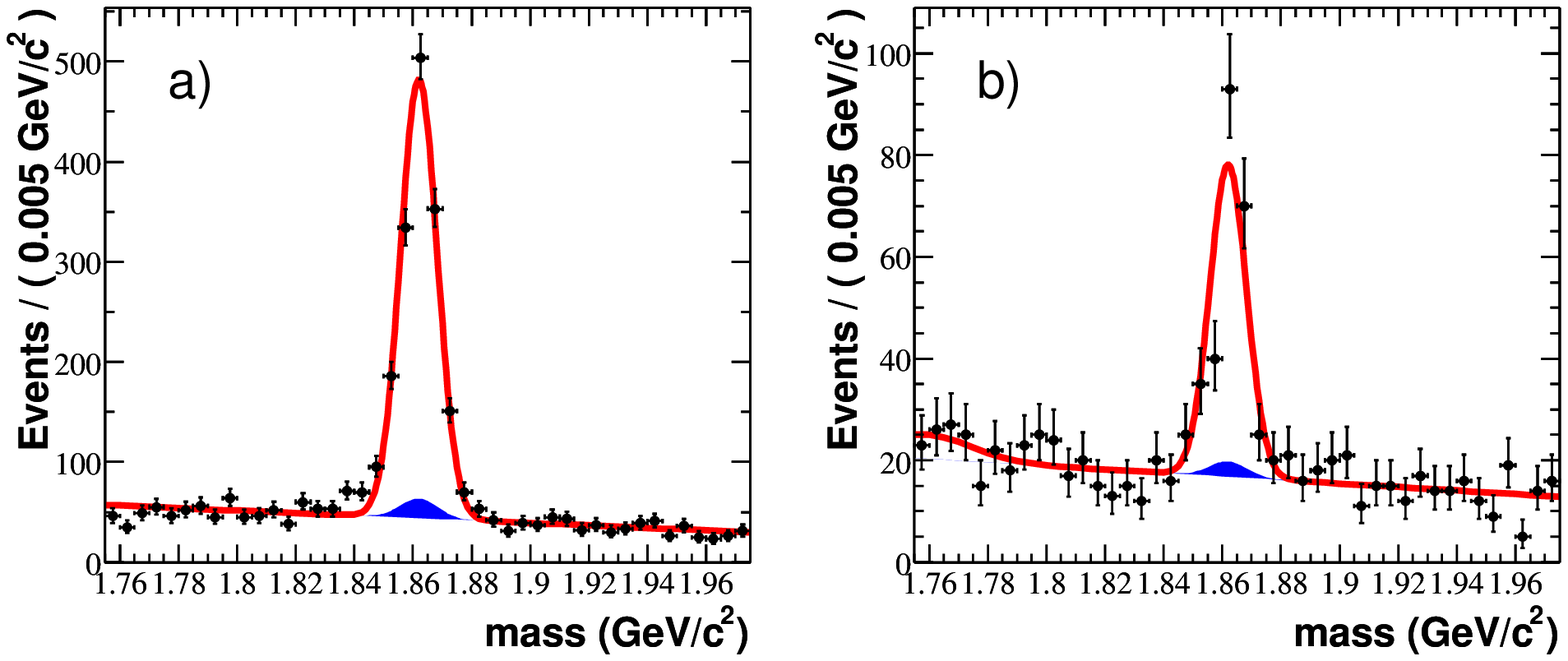}
\includegraphics[width=0.98\linewidth,height=0.42\linewidth]{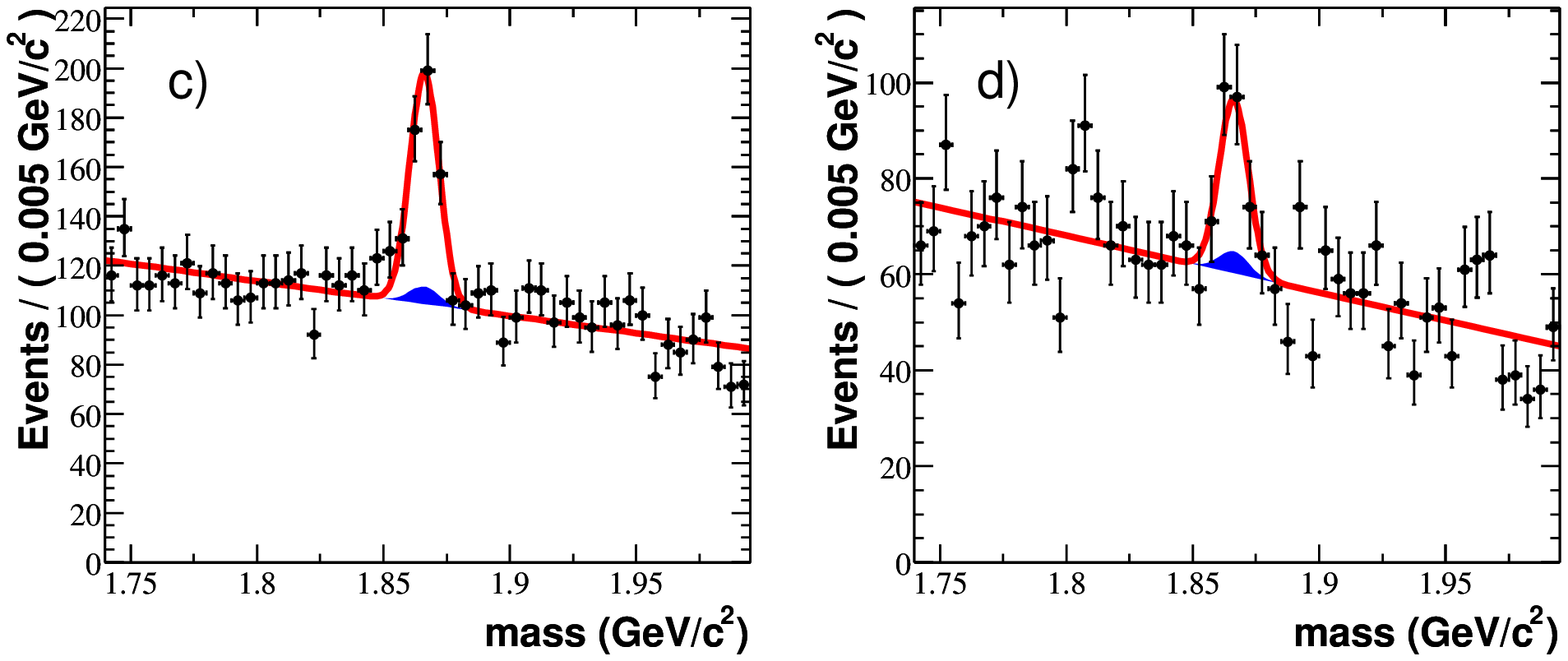}
\includegraphics[width=0.98\linewidth,height=0.42\linewidth]{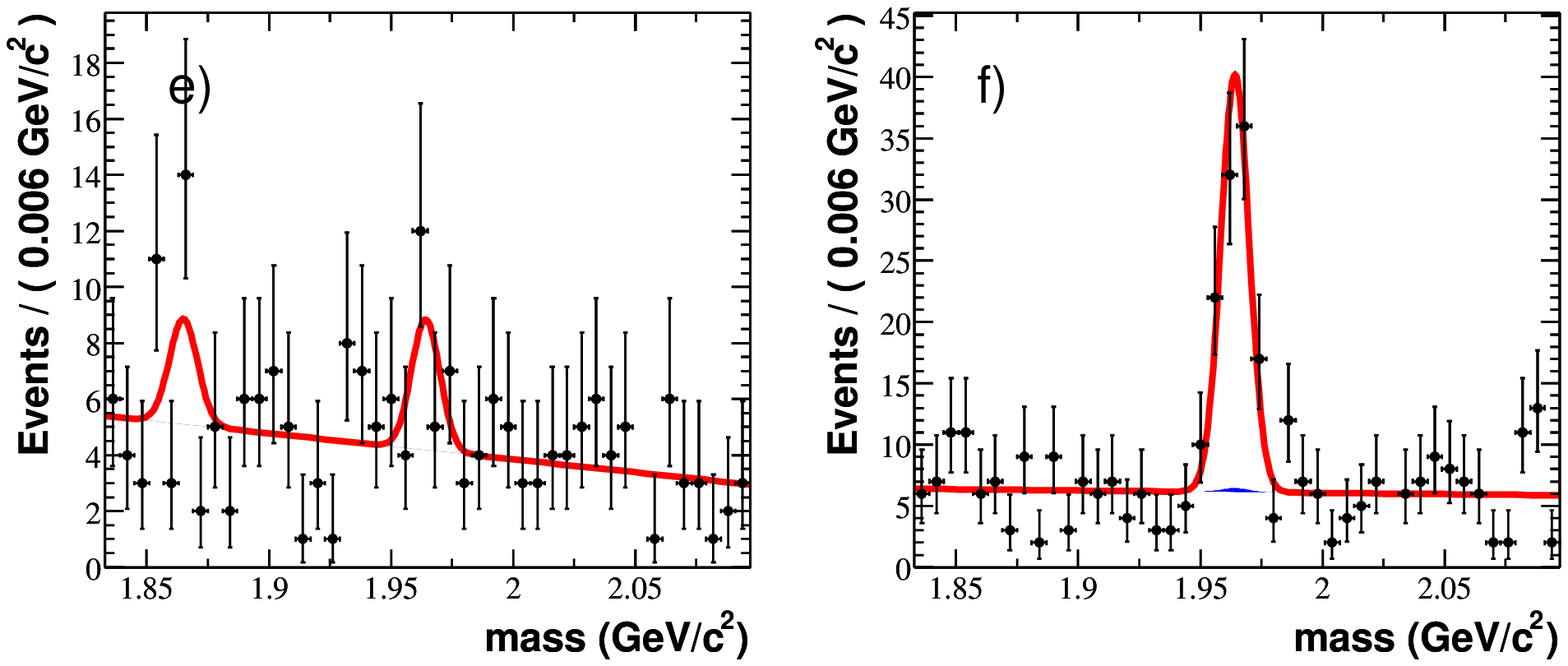}
\includegraphics[width=0.98\linewidth,height=0.42\linewidth]{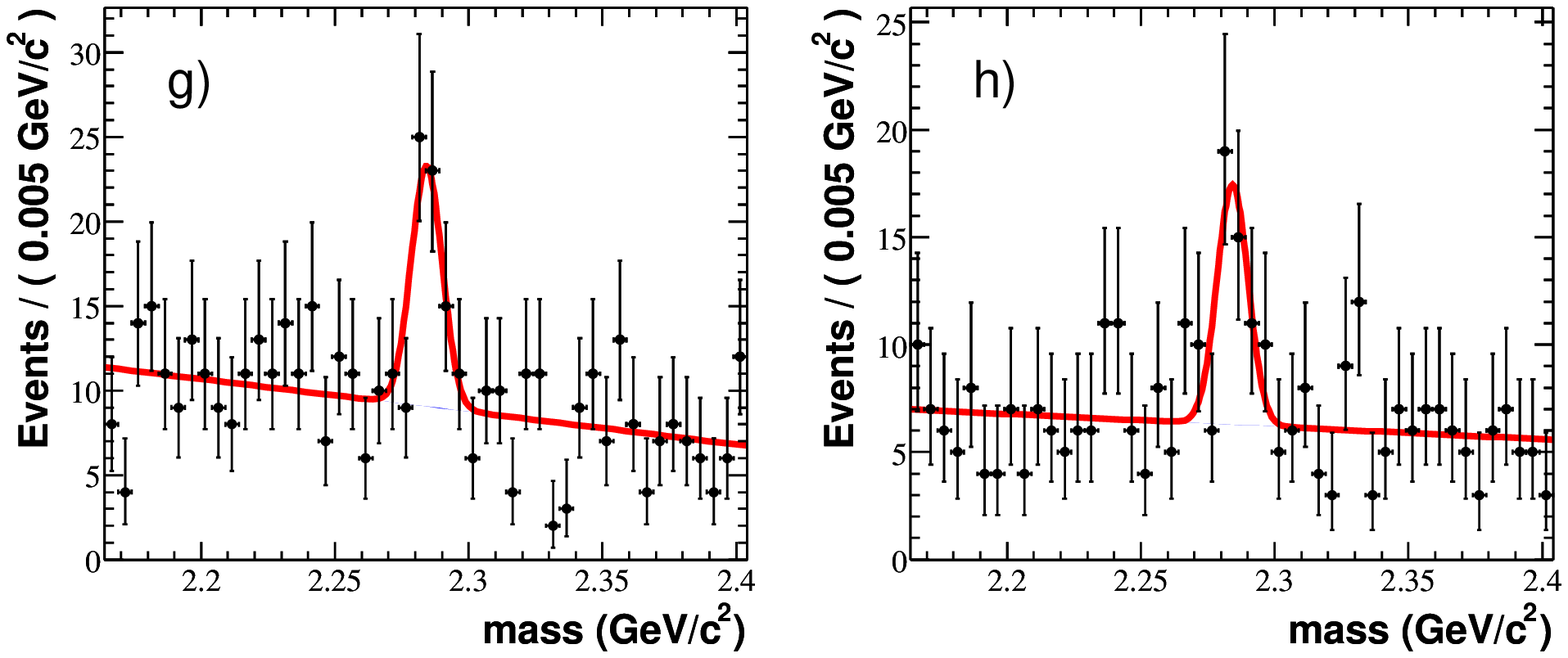}
\caption{Correlated (left) and anti-correlated (right) charmed
particle mass spectra in the recoil of \Bp events, for (a),(b)
$\Dz\to \Km\pip$; (c),(d) $\Dp\to \Km\pip\pip$; (e),(f) $\Dsp\to
\phi\pip$; and (g),(h) $\Lcp\to \proton\Km\pip$. The solid curve
is the sum of a Gaussian signal and of a linear background plus
mode-dependent satellite contributions~\cite{satellite_contributions}. The shaded areas show the
contribution of well reconstructed $D$, \Dsr or \Lc in the \Bp
combinatorial background.} \label{fig:dmass}
\end{center}
\end{figure}

For charged \B decays, Fig.~\ref{fig:dmass} shows the $D$, \Dsr, and \Lc mass spectra 
of correlated and anti-correlated candidates recoiling against $B$'s reconstructed 
in the \mes signal region, for some selected decay modes. 
These spectra are fitted with the sum of a Gaussian signal and a linear background 
(including a satellite peak for some channels~\cite{satellite_contributions}). 
The shaded areas correspond to well reconstructed $D$, \Dsr or \Lc 
from the combinatorial \Brec background. 
They are obtained from data in the \mes background control region, 
normalized to the number of combinatorial background events expected under the \Brec peak.  
The background-subtracted reconstructed signal yields are listed in Table~\ref{table:t1}. 
The reconstruction efficiencies for each charmed (anti-charmed) final state 
$\X_c\to f$ ($\Xb_c\to \fbar$) are computed from the simulation 
as a function of the charmed-particle momentum in the \Bm center-of-mass frame, 
and are applied event-by-event to obtain the efficiency-corrected charm signal yields 
$N(X_\c\to f)$ ($N(\Xb_c\to \fbar)$). 
The final branching fractions are computed from these yields, 
the number of \Brec, and the intermediate branching fractions $\BR(\X_c\to f)$
taken from~\cite{PDG2002}. 
They are given by
\begin{eqnarray}
  \BR(\Bm \to \X_c X) = \frac{N(\X_c\to f)}{N_{\Bp}\times \BR(\X_c\to f)}- c_0\BR_0.
\end{eqnarray}

\noindent
Here the raw branching fraction for $\Bm\to \X_c \X$ is modified by a small corrective term,
$c_0\BR_0$, that accounts for the \Bz contamination in the reconstructed \Bp sample. 
The factor $\BR_0$ depends on the measured $\Bzb\to \X_c \X$ and $\Bz\to \X_c \X$ branching fractions, 
and on the $\Bz - \Bzb$ mixing parameter $\chi_d$~\cite{PDG2002}.
It ranges from less than $3$\% for \Lc to as much as $50$\% for correlated \Dz and \Dp.
Doubly Cabibbo-suppressed \Dz decays are also taken into account. 
The branching fractions and their errors are given in Table~\ref{table:t1}.
The statistical and systematic uncertainties are computed separately for each channel. 
For example, the $3.9\%$ absolute systematic uncertainty on $\BR(\Bm \ra \Dz(\Km\pip)X)$
reflects the quadratic sum of 
$1.8\%$ attributed to $N_{\Bp}$, 
$1.3\%$ to the error on the rate of true $D$'s in the \B combinatorial background, 
$0.8\%$ to the Monte Carlo statistics, 
$1.2\%$ to the track-finding efficiency, 
$2.5\%$ to the particle identification, 
$1.2\%$ to $c_0$, and $0.1\%$ to $\BR_0$. 
We combine the results from the different \Dz and \Dsr decay modes 
to extract the final branching fractions listed in Table~\ref{table:t2}.

\begin{table}[!htb]
\begin{center}
\caption{Combined \Bm branching fractions. The first uncertainty is statistical, 
the second is systematic, and the third reflects charm branching-fraction 
uncertainties~\cite{PDG2002}.}
\label{table:t2}
\begin{ruledtabular}
\begin{tabular}{lcc}
	& correlated & anti-correlated \\[1mm]
$\X_c$  & $\BR(\Bm\to \X_c  \X)$(\%) 
	& $\BR(\Bm\to \Xb_c \X)$(\%) \\
\hline \\[-3mm]
$\Dz$   &$79.3 \pm 2.5 \pm 4.0^{+2.0}_{-1.9}$ &  $9.8 \pm 0.9 \pm 0.5^{+0.3}_{-0.3}$  \\[1mm]
$\Dp$   & $9.8 \pm 1.2 \pm 1.2^{+0.8}_{-0.7}$ &  $3.8 \pm 0.9 \pm 0.4^{+0.3}_{-0.3}$  \\[1mm]
$\Dsp$  & $0.5 \pm 0.6 \pm 0.2^{+0.2}_{-0.1}$ & $14.3 \pm 1.6 \pm 1.5^{+4.9}_{-3.0}$  \\[1mm]
                     & $< 2.2$ at $90\%$ CL &   \\[1mm]
$\Lcp$  & $3.5 \pm 0.8 \pm 0.3^{+1.3}_{-0.8}$ &  $2.9 \pm 0.8 \pm 0.3^{+1.1}_{-0.6}$  \\
\end{tabular}
\end{ruledtabular}
\end{center}
\end{table}

To extract $N_\c$ from these numbers, we need to evaluate the contribution of $\Bm\to \Lcp \Lcm \Km$. 
Combining the four-momenta of the recoiling $B^-$, of a $K^-$ and of the reconstructed $\Lcp$ or $\Lcm$ candidate, 
we compute the missing mass: the absence of signal at the \Lc mass excludes 
a significant contribution of this process. We therefore take ${\BR(\Bm\to \Xi_c\X) = \BR(\Bm\to\Lcm \X)}$ 
in the computation of $N_\c$. Using Eqs.~\ref{eq:nc} and \ref{eq:ncbar} and taking 
${\BR(\Bm\to (\ccbar) X)}$ = $(2.3 \pm 0.3)\%$~\cite{nclep}~\cite{unkcc}, one obtains:
\begin{eqnarray}
\nonumber
  N_\c^- &=& 0.983 \pm 0.030 \pm 0.046 ^{+0.028}_{-0.023}, \\
\nonumber
  N_{\cbar}^- &=& 0.330 \pm 0.022 \pm 0.020 ^{+0.051}_{-0.031},\\
\nonumber n_\c^- &=& 1.313 \pm 0.037 \pm 0.062 ^{+0.063}_{-0.042}.
\end{eqnarray}

The reconstruction of $D$, \Dsr and \Lc from \Bzb decays is performed in the same way as 
that in the \Bm analysis. The corresponding yields are listed in Table~\ref{table:t1}. We then compute for each decay channel $\X_c\to f$ the efficiency-corrected signal yields $N(\X_c\to f)$ ($N(\Xb_c\to \fbar)$) and define the raw branching fractions $\BR_c$ and $\bar\BR_c$ as
\begin{eqnarray}
  \BR_c            &=& N(\X_c\to f) / [N_{\Bz}\times{\BR(\X_c\to f)}] \\
  \overline{\BR}_c &=& N(\Xb_c\to \fbar) / [N_{\Bz}\times{\BR(\X_c\to f)}].
\end{eqnarray}

After correcting these numbers for \BzBzb mixing, we obtain the final branching fraction for $\Bzb\to \X_c \X$:

\begin{eqnarray}
  \BR(\Bzb\to \X_c \X) &=& \frac{\BR_c-\chi_d(\BR_c+\overline{\BR}_c)- c_+\BR_+}{1-2\chi_d},
\end{eqnarray}
where $\chi_d=0.181\pm0.004$ is the $\Bz - \Bzb$ mixing parameter~\cite{PDG2002}.
The correcting factor $\BR_+$ accounts for \Bp contamination in the \Bz sample 
and depends on $\BR(\Bm\to \X_c\X)$ and $\BR(\Bp\to\X_c\X)$. 
The results are given in Table~\ref{table:t1}. 
Combining the different \Dz or \Dsr modes, 
we obtain the final branching fractions listed in Table~\ref{table:t4}.

\begin{table}[!htb]
\begin{center}
\caption{Combined \Bzb branching fractions. The first uncertainty is statistical, the second is systematic,
and the third reflects charm branching-fraction uncertainties~\cite{PDG2002}.} \label{table:t4}
\begin{ruledtabular}
\begin{tabular}{lcc}
        & correlated & anti-correlated \\[1mm]
$\X_c$  & $\BR(\Bzb \to \X_c \X)$(\%) 
        & $\BR(\Bzb \to \Xb_c\X)$(\%) \\
\hline \\[-3mm]
$\Dz$  &      $51.1 \pm 3.1 \pm 2.5^{+1.3}_{-1.3}$ &  $6.3 \pm 1.9 \pm 0.5^{+0.2}_{-0.2}$ \\[1mm]
$\Dp$  &      $39.7 \pm 3.0 \pm 2.8^{+2.8}_{-2.5}$ &  $2.3 \pm 1.8 \pm 0.3^{+0.2}_{-0.2}$ \\[1mm]
                     && $< 5.1$ at $90\%$ CL    \\[1mm]
$\Dsp$ &      $3.9 \pm 1.7 \pm 0.4^{+1.3}_{-0.8}$ & $10.9 \pm 2.1 \pm 0.8^{+3.8}_{-2.3}$ \\[1mm]
                     & $< 8.7$ at $90\%$ CL &   \\[1mm]
$\Lcp$ &      $4.9 \pm 1.7 \pm 0.4^{+1.8}_{-1.0}$ &  $2.0 \pm 1.2 \pm 0.2^{+0.7}_{-0.4}$ \\[1mm]
                     && $< 3.8$ at $90\%$ CL    \\
\end{tabular}
\end{ruledtabular}
\end{center}
\end{table}

To compute $N_\c$, we neglect $\Bzb\to\Lcp \Lcm \Kz$ production and assume that 
${\BR(\Bzb\to \Xi_c \X) = \BR(\Bzb\to \Lcm X)}$. 
Substituting \Bzb for \Bm in Eqs.~\ref{eq:nc} and \ref{eq:ncbar} and taking 
${\BR(\Bz\to (\ccbar) X)}$ = $(2.3 \pm 0.3)\%$~\cite{nclep}~\cite{unkcc}, we obtain:
\begin{eqnarray}
\nonumber
N_\c^0      &=& 1.039 \pm 0.051 \pm 0.049 ^{+0.039}_{-0.031},\\
\nonumber 
N_{\cbar}^0 &=& 0.237 \pm 0.036 \pm 0.012^{+0.039}_{-0.024}, \\
\nonumber 
n_\c^0      &=& 1.276 \pm 0.062 \pm 0.058 ^{+0.066}_{-0.046}.
\end{eqnarray}

We also compute the fraction of anti-correlated charm production in \B decays, 
$w(\Xb_c)=\BR(\Bb\to \Xb_c X)
/ (\BR(\Bb\to \X_c \X)+\BR(\Bb\to \Xb_c \X))$. Here, many systematic uncertainties cancel (tracking, 
\kaon identification, $D$ branching fractions, \B counting). The results are given in Table~\ref{table:t5}. 
We obtain an upper limit on the correlated $\Dsp$ fraction in \Bm decays~:
$ \BR(\Bm\to \Dsp X) / \BR(\Bm\to \Dspm X)<0.126$ at $90\%$ CL.

\begin{table}[!htb]
\begin{center}
\caption{Fraction $w$ of anti-correlated charm.} \label{table:t5}
\begin{ruledtabular}
\begin{tabular}{lcc}
Mode  & $\Bm$ decays & $\Bzb$ decays \\
\hline \\[-3mm]
$\Dzb X$  & $0.110\pm0.010\pm0.003$ & $0.110\pm0.031\pm0.008$ \\[1mm]
$\Dm X$   & $0.278\pm0.052\pm0.009$ & $0.055\pm0.040\pm0.006$ \\[1mm]
$\Dsm X$  & $0.966\pm0.039\pm0.012$ & $0.733\pm0.092\pm0.010$ \\[1mm]
$\Lcm X$  & $0.452\pm0.090\pm0.003$ & $0.286\pm0.142\pm0.007$ \\
\end{tabular}
\end{ruledtabular}
\end{center}
\end{table}

In conclusion, we have measured for the first time the branching fractions for inclusive decays of $B$ mesons to flavor-tagged $D$, \Dsr and \Lc, separately for \Bm and \Bzb. We observe significant production of anti-correlated \Dz and \Dp mesons in \B decays (Table~\ref{table:t5}), with the branching fractions detailed in Tables~\ref{table:t2} and \ref{table:t4}. The correlated \Dsr production in \Bm decays is measured to be small.

As expected, the sum of all correlated charm branching fractions, $N_\c$, is compatible with 1, for charged as well as for neutral \B's. The numbers of charmed particles per \Bm decay  ($n_\c^- = 1.313 \pm 0.037 \pm 0.062 ^{+0.063}_{-0.042} $) and per \Bzb decay ($n_\c^0 = 1.276\pm 0.062\pm 0.058^{+0.066}_{-0.046}$) are consistent with previous measurements~\cite{cleoinc,cleonc,nclep} and with theoretical expectations~\cite{ncbagan,ncbuchalla,ncneubert}.

We are grateful for the excellent luminosity and machine conditions
provided by our \pep2\ colleagues, 
and for the substantial dedicated effort from
the computing organizations that support \babar.
The collaborating institutions wish to thank 
SLAC for its support and kind hospitality. 
This work is supported by
DOE
and NSF (USA),
NSERC (Canada),
IHEP (China),
CEA and
CNRS-IN2P3
(France),
BMBF and DFG
(Germany),
INFN (Italy),
FOM (The Netherlands),
NFR (Norway),
MIST (Russia), and
PPARC (United Kingdom). 
Individuals have received support from CONACyT (Mexico), A.~P.~Sloan Foundation, 
Research Corporation,
and Alexander von Humboldt Foundation.


\begin{thebibliography}{99}
\bibitem{chconj} Throughout this paper, the named reaction refers also to its complex conjugate.
\end{thebibliography}
\end{document}